\documentclass[letterpaper,12pt]{article}   
\usepackage{osajnl2} 
\usepackage[draft]{hyperref} 
\usepackage{jnl_macros}
\usepackage{amsmath}
\usepackage{amssymb}
\usepackage{multirow}

\newcommand{\farcs}{\mbox{\ensuremath{.\!\!^{\prime\prime}}}}

\begin{document}

\title{A low cost scheme for high precision dual-wavelength laser metrology}


\author{Yitping Kok,$^{1,*}$ Michael J. Ireland,$^{2,3}$, J. Gordon Robertson$^1$,
Peter G. Tuthill$^1$, Benjamin A. Warrington$^2$ and William J.
Tango$^1$}
\address{$^1$Sydney Institute for Astronomy, School of Physics, \\ University of
Sydney, NSW 2006, Australia}
\address{$^2$Department of Physics and Astronomy, \\ Macquarie University, NSW
2109, Australia}
\address{$^3$Australian Astronomical Observatory, \\ PO Box 915, North Ryde, NSW
1670, Australia}
\address{$^*$Corresponding author: y.kok@physics.usyd.edu.au}

\begin{abstract}
A novel method capable of delivering relative optical path length metrology with
nanometer precision is demonstrated.
Unlike conventional dual-wavelength metrology which employs heterodyne
detection, the method developed in this work utilizes direct detection of
interference
fringes of two He-Ne lasers as well as a less precise stepper motor open-loop
position control system to perform its measurement.
Although the method may be applicable to a variety of
circumstances, the specific application where this metrology is
essential is in an astrometric optical long baseline stellar interferometer
dedicated to precise measurement of stellar positions. In our example
application of this metrology to a narrow-angle astrometric interferometer,
measurement of nanometer precision could be achieved without frequency-stabilized
lasers although the use of such lasers would extend the range of optical path
length the metrology can accurately measure.
Implementation of the method requires very little additional optics or
electronics, thus minimizing cost and effort of implementation.
Furthermore, the optical path traversed by the metrology lasers is identical
with that of the starlight or science beams, even down to using the same
photodetectors, thereby minimizing the non-common-path between
metrology and science channels.


\end{abstract}


\maketitle 

\section{Introduction}

The specific problem motivating this research was to measure the relative length
of optical paths within an astrometric stellar interferometer to a high degree
of precision. Precision in the order of several nanometers allows stellar
interferometers of about 100m baseline\footnote{separation between two
telescopes} to achieve angular
precision of several microarcseconds. One potential science goal that can be
pursued with such angular precision is the detection of exoplanets through
narrow-angle astrometry
\cite{Shao:1988,Armstrong:1998,Colavita:1999,Schuhler:2006a,Gillessen:2012}.
The basic principle of searching for exoplanets through narrow angle astrometry
is to measure the position of a target star with respect to the position of a
reference star on the celestial sphere to an angular precision of tens of
microarcseconds. Through an optical long baseline
stellar interferometer, light from a resolved pair of stars forms interference
fringes at different optical delays and the difference in optical delay is
proportional to the projected separation of the two stars on the celestial
sphere. For these reasons, the relative position of one star with respect to the
other can be determined by measuring the difference in optical path where the
stellar fringes are found.

High precision measurement of an optical path length can be
conducted by analyzing the interference fringes formed by light (e.g.\ from a
laser) traversing the optical path being measured. The maximum optical length
measurable by interferometry is limited to the coherence length of the light
source because interference fringes are not visible beyond this length.
Therefore, lasers, which typically have a long coherence length,
are usually the preferred light source. However due to this same reason and the
periodicity of the interference fringes, measurement of an optical path length
can be ambiguous. The fringe patterns formed by two optical path lengths that
differ by exactly one laser wavelength are exactly the same. The measurement is
unambiguous only if the optical path length to be measured is within one laser
wavelength. This limited range of distance where the metrology can measure
accurately is also known as the non-ambiguity range (NAR).
A straightforward solution to measure
distances larger than the NAR of a single wavelength metrology is to have the
optical path length be incremented from zero to the desired length in steps no
larger than the NAR and incremental measurement is
carried out at each step. But due to practical requirements this solution is not
always desirable. Conventionally, instead of one wavelength, two laser
wavelengths are used to obtain a long synthetic optical wavelength by means of
heterodyne interferometry \cite{Daendliker:1988,Schuhler:2006}. However, instead
of using the heterodyne detection technique which requires specialized optical
elements and hence has higher cost, the dual-wavelength metrology described
here and implemented at the Sydney University Stellar Interferometer (SUSI)
\cite{Davis:1999} employs a simple homodyne fringe counting detection scheme
together with a (relatively) less precise stepper motor open-loop
position control system to extend the range of distance the metrology can
accurately measure. The implementation and performance of this metrology, which
was found to be easily suitable for our demanding narrow-angle astrometric
application with an optical long baseline stellar interferometer, are described
hereafter.

\section{Optical setup}
\begin{figure}
\centering
\includegraphics[width=0.9\textwidth]{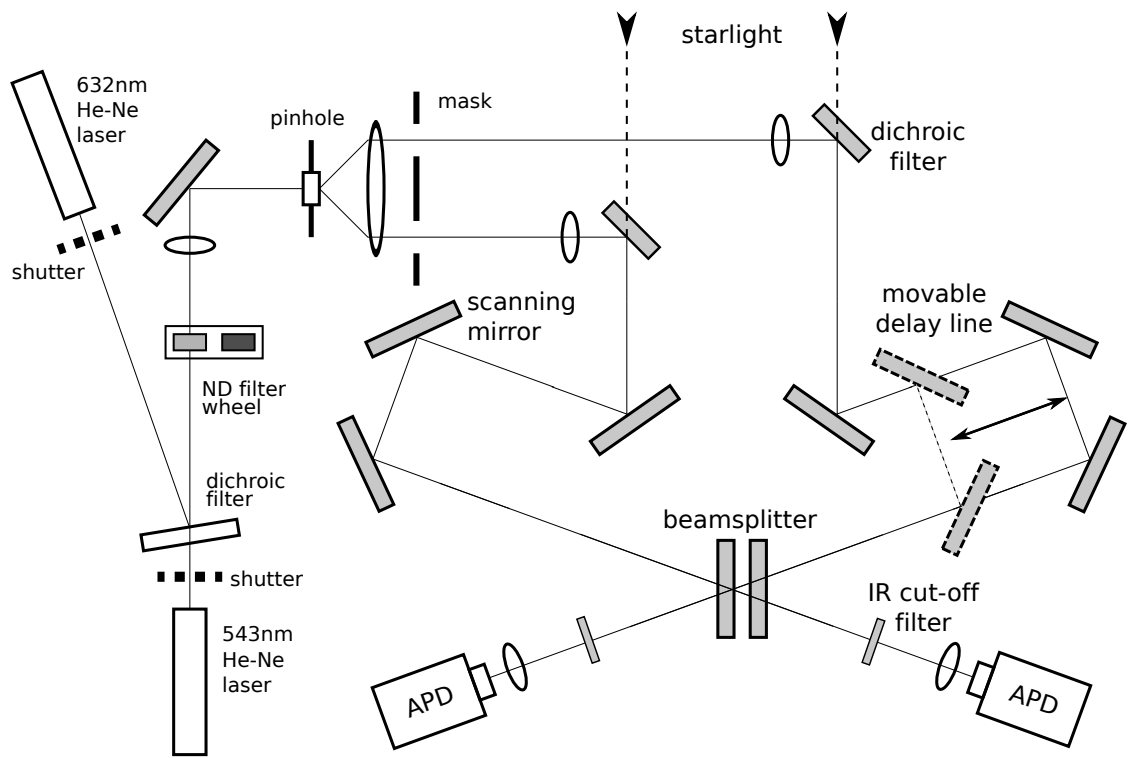}
\caption{Optical setup of the dual-wavelength metrology and the astrometric beam
combiner in SUSI. f1.eps.}
\label{fig:optics}
\end{figure}

The diagram in Fig.~\ref{fig:optics} shows a simplified version of the
narrow-angle astrometric beam combiner (MUSCA) in SUSI. Instead of depicting the
entire SUSI facility for which schematic diagrams can be obtained from
\cite{Robertson:2012}, the diagram shows only the optical path relevant to the
metrology and the beam combiner which is a pupil-plane Michelson interferometer.
%
%
The light sources for the metrology are two He-Ne lasers; one emits at peak
wavelength of $\lambda_r=\sigma_r^{-1}=0.6329915\mu$m and the other at peak
wavelength of $\lambda_g=\sigma_g^{-1}=0.5435161\mu$m
\cite[converted from wavelengths in standard (760 torr, 15$^\circ$C) dry
air]{Kurucz:2012,Edlen:1966}. The quoted values are wavelengths in vacuum but
the two lasers are operated in air. Both laser beams are first spatially
filtered by pinholes, then collimated and finally refocused into the
interferometer. Each refocused beam forms an image at a field lens in front of
an avalanche photodiode (APD).
The optical path along the left arm of the interferometer (as seen in
Fig.~\ref{fig:optics}) is periodically modulated by a piezo-electrically
actuated mirror (scanning mirror) to produce temporal fringes, which are then
recorded by the pair of APDs as a time series of photon counts. The scanning
mirror modulates the optical path in 256 discrete steps in about 70ms per scan
period per scan direction. On the right arm of the interferometer, the length of
the optical path can be changed by a movable delay line. It is made up of
mirrors sitting on a linear translation stage
(Zaber\footnote{\url{http://www.zaber.com}} T-LS28M) which is stepper motor
driven and has an open loop position control system. The built-in stepper motor
converts rotary motion to linear motion via a leadscrew. The leadscrew based
open loop position control system has a nominal accuracy of 15$\mu$m.

It is important to note that, apart from the lasers and their injection optics,
all components in Fig.~\ref{fig:optics} were pre-existing and required for the
science goal of the beam combiner. During astronomical observations, the same
pair of APDs are used to record both the stellar and the metrology fringes. The
optical path of the metrology lasers is designed to trace the optical path of
the starlight beams in the beam combiner, which propagates into
the instrument from the top as seen in Fig.~\ref{fig:optics} through a pair of
dichroic filters. In this way, the optical path probed by the metrology is
nearly identical to the optical path of the starlight and the small difference
in optical delay (due to the difference in wavelengths) is invariant under the
controlled atmospheric condition \cite{Erickson:1962} in the laboratory in which
the optics are housed.

\section{Dual-wavelength metrology} \label{sec:principle}
Our dual-wavelength metrology is designed to measure the change in optical path
length of air brought about by a displacement of the delay line
when it is moved from one position to another.
The underlying principle of the metrology is to first measure phases of
interference fringes of two lasers, operating at wavelengths whose ratio is
theoretically not a rational number (but practically a ratio of two large
integers due to finite accuracy of the wavelengths), at two different
delay line positions and then determine the
number of fringe cycles that have evolved as a result of the displacement.
The phase measurement is key in this classical two-wavelength approach to
displacement metrology. The novel aspect of the metrology described here is the
use of optical path modulation to measure the phases of the fringes of the two
lasers simultaneously.
In an idealized case where the laser wavelengths are perfectly stable and the
measurements of the phases are noiseless, one measurement at each delay line
position would be enough to uniquely resolve the length of the optical
path between the two positions.
However, in the real world, due to uncertainties in the phase measurements and
laser wavelengths, there are a series of plausible solutions for the optical
path length. The span between these plausible solutions is the non-ambiguity
range (NAR) of the metrology and is elaborated in Section~\ref{sec:nar}.
In order to extend the range of distance the metrology can measure an open loop
stepper motor position control system is exploited to narrow down the plausible
solutions to a single best fit, thereby yielding the displaced optical path length
measurement at interferometric precision.
The basic requirement for this two-prong approach is that the NAR arising from
fringe phase measurement must be larger than the uncertainty of the stepper
motor positioning system.
Since the stepper motor positioning system can determine the position of the
delay line unambiguously over a large distance range (in the case of T-LS28M,
28mm), the delay line can be moved quickly ($\sim$2mm/s) from one position to
another and fringe phase measurement does not have to be done on the
fly but before and after a move.


In order to explain the method in more detail, first, let the distance between a
position of the delay line and an arbitrary reference position be expressed in
terms of two laser wavenumbers ($\sigma_r$ and $\sigma_g$) as follows,
\begin{equation} \label{eq:met_di}
d_i = \frac{1}{n_r\sigma_r}\left(N_{ri} + \frac{\phi_{ri}}{2\pi}\right) =
\frac{1}{n_g\sigma_g}\left(N_{gi} + \frac{\phi_{gi}}{2\pi}\right)
\end{equation}
$n_r$ and $n_g$ are the refractive indices of air at the respective wavenumbers
while $N_{ri}$ and $\phi_{ri}/2\pi$ or $N_{gi}$ and $\phi_{gi}/2\pi$ are the
number of full (integer) and fractional wavelengths that fit within this
distance. The subscript $i$ represents one position of the delay line and if two
different positions are considered, then, from Eq.~\eqref{eq:met_di},
\begin{equation} \label{eq:met_d}
\begin{split}
d &= d_1 - d_0 \\
  &= \frac{\left(\Delta\varphi_1-\Delta\varphi_0\right)/2\pi + \Delta N}{n_r\sigma_r - n_g\sigma_g} \\
\end{split}
\end{equation}
where $d$ represents the difference in optical path length of air between the
two delay line positions while $\Delta\varphi_i = \phi_{ri}-\phi_{gi}$, $\Delta
N = \Delta N_r - \Delta N_g$, $\Delta N_r = N_{r1}-N_{r0}$ and $\Delta N_g =
N_{g1}-N_{g0}$. The phases of the laser fringes, $\phi_{ri}$ and $\phi_{gi}$,
and their difference, $\Delta\varphi_i$, can be obtained from the photon counts
recorded by the APDs. $\Delta N_r$ and $\Delta N_g$, on the other hand, cannot
be directly determined but can be inferred from the following equality,
\begin{equation}
\label{eq:met_ni}
\Delta N_g - \alpha \Delta N_r = (\Delta\psi_1 - \Delta\psi_0)/2\pi
\end{equation}
where $\alpha = \frac{n_g\sigma_g}{n_r\sigma_r}$ and $\Delta\psi_i =
\alpha\phi_{ri} - \phi_{gi}$. Therefore the main observables for the metrology
at each delay line position are $\phi_{ri}$, $\phi_{gi}$, $\Delta\varphi_i$ and
$\Delta\psi_i$.

The values of $\Delta N_g$ and $\Delta N_r$ are determined through a
model-fitting method based on Eq.~\eqref{eq:met_ni}. First, a range of guess
values are generated based on the optical path length estimated from the stepper
motor positioning system, $d_{\rm{zaber}}$, and the NAR to evaluate the LHS of
Eq.~\eqref{eq:met_ni}. Next the result is compared with the RHS of
Eq.\eqref{eq:met_ni} which is obtained from the phase measurement.
Theoretically, there is a unique set of $\Delta N_g$ and $\Delta N_r$ values
that satisfy the equation because $\alpha$ is an irrational number. However, due
to uncertainty in the phase measurement this is not the case in practice.
Instead the set of $\Delta N_g$ and $\Delta N_r$ that minimizes the error
between the RHS and LHS of the equation is the set of values to be used for
distance determination in Eq.~\eqref{eq:met_d}.

\section{Non-ambiguity range} \label{sec:nar}

As previously described, the metrology measures the length of an optical path by
calculating the number of laser wavelengths that
can be fitted into it. However, since the ratio of the laser wavenumbers,
$\alpha=1.164644$, can be approximated by a ratio of two integers, e.g.\
$\alpha\approx\frac{7}{6}$ or $\alpha\approx\frac{92}{79}$, the phases of the
laser fringes will appear (depending on the uncertainty of the phase
measurement) to realign after several wavelengths as suggested by the numerator
and denominator of the fraction. This means that the phase differences between
the laser fringes will repeat and become indistinguishable from the previous
phase realignment if the optical path length is larger than the distance
suggested by the wavelength range. Therefore the metrology can only determine
the accurate length of the optical path if it is within this range, which is the
non-ambiguity range (NAR) of the metrology.


The parameters, $\Delta N_r$ and $\Delta N_g$, can take any integer value but in
order to determine the value of NAR of the metrology, suppose $0 \le \Delta N_r
\le \Delta N_{r,\rm{max}}$ and $0 \le \Delta N_g \le \Delta N_{g,\rm{max}}$.
Then the NAR is defined as,
\begin{equation}
\text{NAR} = \min
  \left(
  \Delta N_{r,\rm{max}}/\sigma_r,\; \Delta N_{g,\rm{max}}/\sigma_g
  \right)
\end{equation}
provided that the following inequalities are satisfied for all values of $\Delta
N_r$ and $\Delta N_g$,
\begin{equation}
\begin{split}
|\alpha\Delta N_r - [\alpha\Delta N_r]| &> \delta(\Delta\psi)/2\pi \\
|\Delta N_g/\alpha - [\Delta N_g/\alpha]| &> \delta(\Delta\psi)/2\pi
\end{split}\end{equation}
which is derived from the LHS of Eq.~\eqref{eq:met_ni} when the phases of the
lasers fringes are aligned, hence the RHS is zero. The notation $[\cdot]$
denotes the nearest integer of the real number within the brackets and
$\delta(\Delta\psi)$ is the standard error of mean of
$\Delta\psi = \Delta\psi_1 - \Delta\psi_0$.


\section{Implementation}
\begin{figure}
\centering
\includegraphics[width=0.9\textwidth]{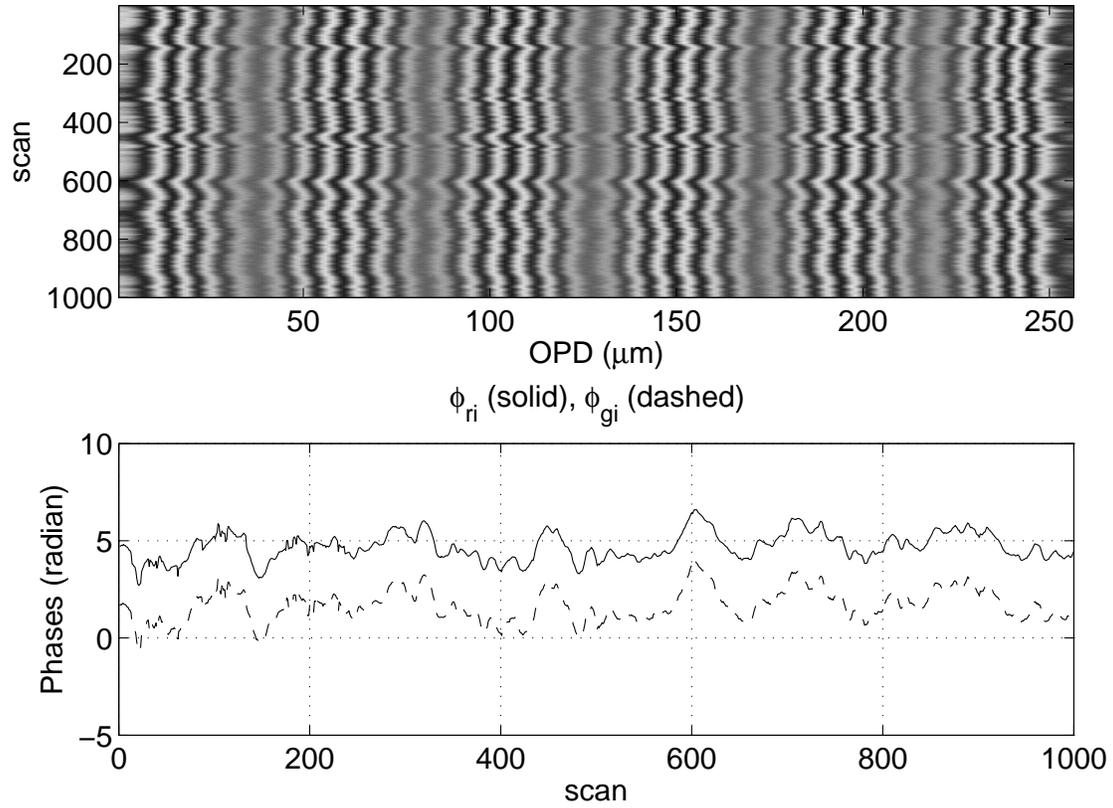}
\caption{Consecutive scans of laser fringes (top) and the phases (bottom), $\phi_{ri}$
and $\phi_{gi}$, extracted from each scan. The internal laboratory seeing was poor at
the time of measurement. f2.eps.}
\label{fig:phi}
\end{figure}

The photon counts recorded from the setup in Fig.~\ref{fig:optics} are reduced
with a program written in
MATLAB\footnote{\url{http://www.mathworks.com/products/matlab}}/Octave\footnote{\url{http://www.gnu.org/software/octave}}
to determine optical path differences based on the model presented in the
previous sections. For each set of laser fringes recorded at one position of the
delay line the phases of the fringes (relative to the middle of the scan
length),
namely $\phi_{ri}$, $\phi_{gi}$, $\Delta\varphi_i$ and $\Delta\psi_i$, 
are extracted using a Fast Fourier Transform (FFT) routine. Fig.~\ref{fig:phi}
shows the laser fringes and the phases extracted from them. Because the
implementation is numerical and the FFT routine expresses phases in the range of
$-\pi$ to $\pi$, a minor tweak to the value of $\Delta N$ in
Eq.~\eqref{eq:met_d} may be required to obtained an accurate value of $d$. As a
result, the term should be replaced with $\Delta N'$, where,
\begin{equation} \label{eq:met_deltan}
\Delta N' = \Delta N + \delta(\Delta N)
\end{equation}
The adjustment, $\delta(\Delta N)$, which value is obtained from computer
simulation, consists of two parts and is summarized below as,
\begin{equation} \label{eq:met_deldeltan}
\begin{split}
\delta(\Delta N) &= \delta(\Delta N)_0 \\
&\quad+ \left\{
  \begin{array}{l l}
  -1 & \text{if}\;\Delta\phi_r < 0, \Delta\phi_g \ge 0; \\
  0  & \text{if}\;\Delta\phi_r < 0, \Delta\phi_g < 0  ; \text{or} \\
     & \;\;\;\Delta\phi_r \ge 0, \Delta\phi_g \ge 0; \\
  1  & \text{if}\;\Delta\phi_r \ge 0, \Delta\phi_g < 0. \\
  \end{array}\right.
\end{split}\end{equation}
where $\Delta\phi_r=\phi_{r1}-\phi_{r0}$ and $\Delta\phi_g=\phi_{g1}-\phi_{g0}$.
The value of $\delta(\Delta N)_0$ is given in Table~\ref{tab:met_delN} if all
the expressions in the first five columns in the table are satisfied, otherwise
$\delta(\Delta N)_0$ is zero.
For example, according to the first row of Table~\ref{tab:met_delN}, if
$\Delta\varphi_1 \ge \Delta\varphi_0$, $\Delta\phi_r \ge 0$, $\Delta\phi_g \ge
0$, $|\Delta\phi_r| \le \pi$ and $|\Delta\phi_g > \pi$, then $\delta(\Delta
N)_0 = -1$.
The differences of phases in
Eq.~\eqref{eq:met_deldeltan} and Table~\ref{tab:met_delN} are computed by first
expressing the phases in the range of 0 to $2\pi$.

The measurement of the phases of the lasers fringes are carried out before and
after the delay line is moved for astronomical observation. The displacement of
the delay line brings one of the two stellar fringe
packets\footnote{interference fringes localized in delay space due to limited
coherence length of the source} into the scan range of the scanning mirror. By
measuring the displacement of the delay line and the position of the fringe
packets, the optical delay between them, which is the main science observable of
MUSCA, can be measured. MUSCA spends about 15--30 minutes, depending on seeing
condition of the night sky, integrating on each fringe packet while the
metrology takes about 2--3 minutes in total to measure phases of the lasers
fringes. The time spent by the metrology includes moving the delay line from one
position to another. This sequence of astronomical and metrology measurements is
repeated at least 3 times for each science target.

\section{Sources of error}
The precision of the measurement by the dual-wavelength metrology depends on
several factors which will be elaborated individually in this section.

\subsection{Phase error}
\begin{figure}
\centering
\includegraphics[width=0.9\textwidth]{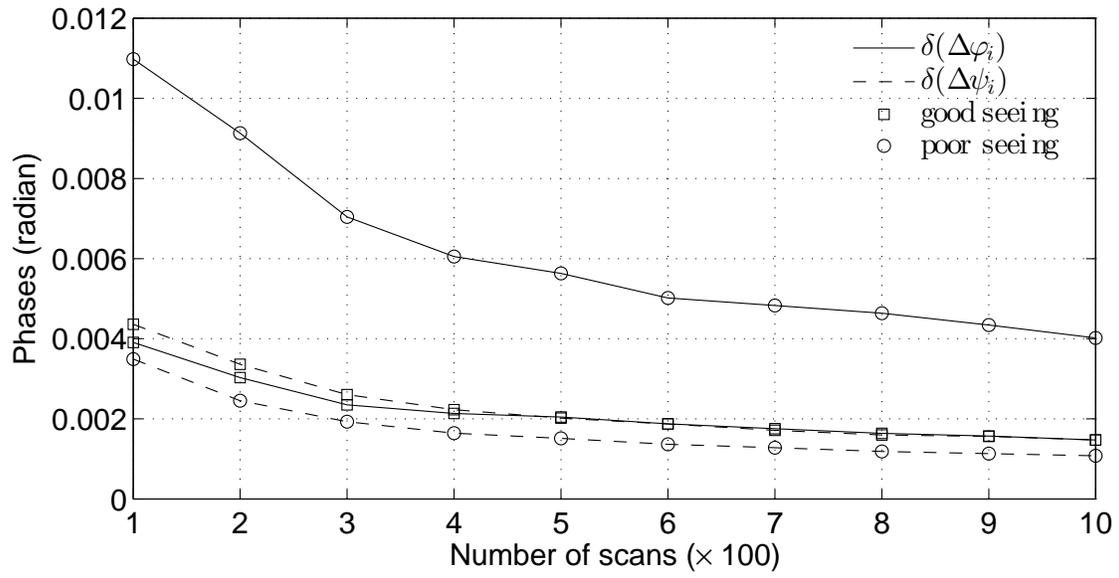}
\caption{The uncertainty of $\Delta\psi_i$ and $\Delta\varphi_i$ measured with different
number of scans of laser fringes. The measurement of the latter is more susceptible to
internal seeing condition in the lab. f3.eps.}
\label{fig:delphi}
\end{figure}

This is the main source of error affecting the precision of the metrology. Errors
in measuring $\Delta\psi_i$ and $\Delta\varphi_i$ determine the uncertainty in
choosing the right value for $\Delta N_r$ and $\Delta N_g$ and the uncertainty
of $d$ respectively.
The physical processes contributing to this error are photon noise and internal
laboratory seeing. At high photon count rates (about 10$^6$ counts per second in
the SUSI setup), the uncertainty of the phase information obtained from a FFT
(or more generally a Discrete Fourier Transform) routine is negligible (i.e.\ in
the order of $10^{-5}$ radians) \cite{Walkup:1973}. Therefore, internal
laboratory seeing is the dominant factor.
Fig.~\ref{fig:delphi} shows the standard
error of the mean of typical measurements of $\Delta\psi_i$ and
$\Delta\varphi_i$. The errors decrease with increasing number of scans. If the
uncertainty of $\Delta\psi_i$ is less than 0.002 radian (with $\gtrsim$500
scans), then the NAR of this metrology is estimated to be $\sim$460$\mu$m
($\Delta N_{r,\rm{max}}=735$).

\subsection{Instrumental error}
\begin{figure}
\centering
\includegraphics[width=0.9\textwidth]{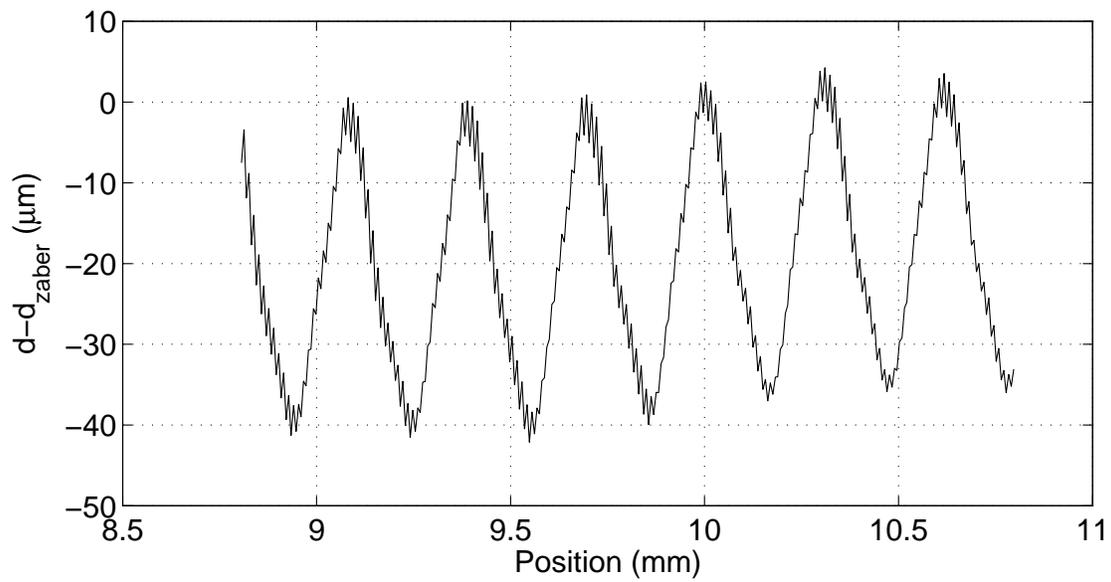}
\caption{The difference between the OPD measured by the dual-laser metrology described
in this section, $d$, and the OPD estimated with the stepper motor metrology of the
Zaber stage, $d_{\rm{zaber}}$, at different stage position. f4.eps.}
\label{fig:char1}
\end{figure}

Given an NAR of $\sim$460$\mu$m the difference between an initial guess optical
path length, $d_{\rm{zaber}}$, and its true value must be less than the NAR
value. The initial guess value is obtained from the stepper motor positioning
system. The characterization of the precision of the system is shown in
Fig.~\ref{fig:char1}. The plot in the figure shows the difference between the
position of the delay line indicated by the stepper motor positioning system and
the position measured by the dual-laser metrology. The cyclical error as seen
from the plot is typical for a leadscrew based linear translation stage
\cite{Zaber:2006}. Being able to reproduce such a cyclical pattern verifies the
accuracy of the dual-laser metrology especially the accuracy of $\Delta N'$.
Instead of the specified 15$\mu$m accuracy Fig.~\ref{fig:char1} shows that the
leadscrew
has a precision of $\sim$20$\mu$m which is still well within the NAR
requirement. This requirement is satisfied even though the optical path length
change induced by the delay line is twice its actual physical change in position
(refer to Fig.~\ref{fig:optics}).

\subsection{Laser wavelength error}
Based on the longitudinal mode spacing specification of the laser (438MHz for
the red and $\sim$350MHz for the green laser) and the theoretical full width
half maximum (FWHM) of the gain profile at the laser wavelengths (1.8GHz for the
red and 1.5GHz for the green laser \cite{Svelto:1998}), the relative uncertainty
of the wavelength, $\frac{\delta\sigma_r}{\sigma_r}$ and
$\frac{\delta\sigma_g}{\sigma_g}$, of individual laser is better than
$3\times10^{-7}$. Here the notation $\delta(\cdot)$ means one standard deviation
of the wavenumber variation. If $\Delta\sigma_{rg} = n_r\sigma_r-n_g\sigma_g$,
then,
\begin{equation}
\frac{\delta(\Delta\sigma_{rg})}{\Delta\sigma_{rg}} \approx 3.6\times10^{-6}
\end{equation}
because the refractive indices are similar ($n_r-n_g \approx 1\times10^{-6}$)
and approximately constant \cite{Erickson:1962} between the time when the laser
fringes are recorded at the two delay line positions. In the case of SUSI, this
condition is true because the fluctuations of ambient temperature in the
laboratory are designed to be small within a typical duration of an astronomical
observation \cite{Davis:1999}. Laser wavelength error of this magnitude is not
significant when measuring short optical path length but can lead to substantial
error in optical path length measurement if the optical path is long. This and
the effect of using frequency-stabilized lasers will be discussed in
Section~\ref{sec:uncertainty}.

\subsection{Non-common-path error}
Other than being used as light sources for the metrology, the lasers are also
used for optical alignment for MUSCA and the rest of the optical setup at SUSI.
In the case of MUSCA the alignment between the lasers and starlight beams is
critical in order to minimize the non-common-path between the metrology and the
science channel. Several other optical elements in the full optical setup at
SUSI which also play a role in assisting the alignment process (e.g.\
retro-reflecting mirrors, lenses, a camera in SUSI's main beam combiner, etc.)
are not included in the simplified version the setup in Fig.~\ref{fig:optics}
but can be referred from \cite{Robertson:2012}. The optics put the pupil and the
image of the pinhole and a star on the same respective planes through the
aperture of the mask. The lasers and starlight beams should ideally be coaxially
aligned in order to minimize the non-common-path between them. However, in the
actual optical setup there can be a maximum misalignment of 0.5mm between the
pinhole and the image of the star over a distance of about 2m. This translates
to a maximum of 0.3 milliradians of misalignment or $\sim3\times10^{-8}$ of
relative metrology error. In absolute terms, this error is negligible ($\ll$1nm)
for short ($<10$mm) optical path length measurement. However, a more precise
alignment is necessary for measurement of longer optical path.

\section{Uncertainty of measurement} \label{sec:uncertainty}
The uncertainty of the optical path length measurement, $\delta d$, can be
derived from Eq.~\eqref{eq:met_d}.
The precision of the stepper motor positioning system (well within the NAR)
ensured that the uncertainty of $\Delta N'$ is always zero.
The characterization of the delay line in Fig.~\ref{fig:char1} verified this in
practice.
Therefore the uncertainty of the optical path length measurement, given below,
depends only on the uncertainty of the phase measurements and the laser
wavenumbers. In order to simplify the equation,
let $\Delta\varphi = \Delta\varphi_1 - \Delta\varphi_0$.
\begin{equation} \label{eq:met_duallaseropderr}
\begin{split}
\delta d &=
  \sqrt{
  \left(\left|\frac{\partial d}{\partial(\Delta\varphi)}\right|
  \delta(\Delta\varphi)\right)^2 + 
  \left(\left|\frac{\partial d}{\partial(\Delta\sigma_{rg})}\right|
  \delta(\Delta\sigma_{rg})\right)^2
  } \\
&=
  \sqrt{
  \left(\frac{\delta(\Delta\varphi)}{2\pi\Delta\sigma_{rg}}\right)^2 +
  \left(\frac{\delta(\Delta\sigma_{rg})}{\Delta\sigma_{rg}}\,d\right)^2
  } \\
\delta d
&\approx \left|\frac{\delta(\Delta\varphi)}{2\pi\Delta\sigma_{rg}}\right|
,\quad (\delta(\Delta\varphi) > 0.001, d < 0.5\text{mm})
\end{split}\end{equation}
If the phase error, $\delta(\Delta\varphi)$, is more than 1 milliradian, which
is typical for this metrology setup then it can be shown that the contribution
of the wavenumber error is negligible at short optical path ($d < 0.5$mm). This
value is similar to the separation of two fringe packets of two stars with a
projected separation of about 0$\farcs$6 in the sky observed with a 160m baseline
interferometer. The plot in Fig.~\ref{fig:delphi} shows that a phase error,
$\delta(\Delta\varphi_i)$, of $\sim$5 milliradians can be achieved with just 100
scans or more than 500 scans in poor internal (laboratory) seeing conditions.
With such magnitude of phase error and according to
Eq.~\eqref{eq:met_duallaseropderr}, the uncertainty of an optical path length
measurement is in the order of 5nm or less.

The range of optical path where the contribution of
$\frac{\delta(\Delta\sigma_{rg})}{\Delta\sigma_{rg}}$ towards the overall error
is negligible can be extended if frequency-stabilized lasers are used. Such
lasers usually have wavelengths accurate up to $1\times10^{-9}\mu$m or
$\sim1\times10^{-9}$ or smaller in relative error and cost about 3--4 times the
price of a regular He-Ne laser\footnote{\url{http://www.thorlab.com},
\url{http://www.newport.com}, \url{http://www.npl.co.uk}}.
At that precision, $\delta d$
is not dependent on the optical path length until about 1m which is well beyond
the required optical delay for narrow-angle astrometry. However, the extension
of the optical path range may also incur other technical cost, which involves
improving the precision of optical alignment, increasing the range and speed of
the delay line. Therefore, an upgrade to the metrology system described here
should take all these factors into consideration.

\section{Conclusion}
A novel, inexpensive dual-wavelength laser metrology system has been presented
and demonstrated to deliver nanometer precision in an experimental
implementation.
The scheme also boasts the significant advantage of propagating the metrology
lasers along an optical path which is identical to the science beam and
recording both signals with the same detectors, thereby eliminating
non-common-path errors. 
Due to much pre-existing common hardware, this scheme was particularly
straightforward to implement within the context of our specific application
(i.e.\ an optical long baseline astrometric stellar interferometer).
However, because it does not require additional specialized optics (e.g. an
acoustic-optoelectronic modulator (AOM)) or electronics (e.g. a digital
phasemeter) and furthermore has relaxed requirements on the accuracy and
stability of the laser wavelengths, it may appeal to similar application within
the optics community, especially for stellar interferometry.

\section*{Acknowledgments}
This research was supported under the Australian Research Council's Discovery
Project funding scheme. Y.K. was supported by the University of Sydney
International Scholarship (USydIS).






\clearpage
\begin{table}
\centering
\caption{Look-up-table for $\delta(\Delta N)_0$ in
Eq.~\eqref{eq:met_deldeltan}. By default, $\delta(\Delta N)_0$ is 0.}
\label{tab:met_delN}
\begin{tabular}{cccccc} \\
\hline
\multirow{2}{*}{$\Delta\varphi_1 < \Delta\varphi_0$} & $\Delta\phi_r$ & $\Delta\phi_g$ &  
$|\Delta\phi_r|$ & $|\Delta\phi_g|$ & \multirow{2}{*}{$\delta(\Delta N)_0$} \\
 & $< 0$ & $< 0$ & $> \pi$ & $> \pi$ & \\
\hline
F & F & F & F & T & -1\\
F & F & T & F & F & -1\\
F & T & F & T & T & -1\\
F & T & T & T & F & -1\\
T & F & F & T & F & 1 \\
T & F & T & T & T & 1 \\
T & T & F & F & F & 1 \\
T & T & T & F & T & 1 \\
\hline
\multicolumn{6}{p{0.5\textwidth}}{\footnotesize
$\text{F}\equiv\text{FALSE}$;
$\text{T}\equiv\text{TRUE}$}
\end{tabular}
\end{table}

\clearpage

%
%
%
%
\listoffigures


\clearpage


\clearpage

\clearpage

\clearpage

\end{document}